\documentclass[aps,twocolumn,prl,showpacs,amsmath,amssymb,10pt]{revtex4-2}
\usepackage{graphicx}
\usepackage{scrextend}
\usepackage{manfnt,pifont}
\usepackage{hyperref}
\usepackage{url}
\usepackage{lipsum}
\usepackage[font=small]{subcaption}
\usepackage[font=small]{caption}

\usepackage[normalem]{ulem}

\usepackage{url}
\usepackage[outline]{contour}
\usepackage{textgreek}

\usepackage{tikz}
\usetikzlibrary{shapes}

\usepackage{CJK}
\usepackage{enumitem}
\setlist[itemize]{leftmargin=*}

\def\tc{\bar{c}}

\newcommand\wh\widehat

\newcommand\ri{{\mathrm{i}}}
\newcommand\rd{{\mathrm{d}}}
\makeatletter
\DeclareFontFamily{OMX}{MnSymbolE}{}
\DeclareSymbolFont{MnLargeSymbols}{OMX}{MnSymbolE}{m}{n}
\SetSymbolFont{MnLargeSymbols}{bold}{OMX}{MnSymbolE}{b}{n}
\DeclareFontShape{OMX}{MnSymbolE}{m}{n}{
    <-6>  MnSymbolE5
   <6-7>  MnSymbolE6
   <7-8>  MnSymbolE7
   <8-9>  MnSymbolE8
   <9-10> MnSymbolE9
  <10-12> MnSymbolE10
  <12->   MnSymbolE12
}{}
\DeclareFontShape{OMX}{MnSymbolE}{b}{n}{
    <-6>  MnSymbolE-Bold5
   <6-7>  MnSymbolE-Bold6
   <7-8>  MnSymbolE-Bold7
   <8-9>  MnSymbolE-Bold8
   <9-10> MnSymbolE-Bold9
  <10-12> MnSymbolE-Bold10
  <12->   MnSymbolE-Bold12
}{}

\let\llangle\@undefined
\let\rrangle\@undefined
\DeclareMathDelimiter{\llangle}{\mathopen}
                     {MnLargeSymbols}{'164}{MnLargeSymbols}{'164}
\DeclareMathDelimiter{\rrangle}{\mathclose}
                     {MnLargeSymbols}{'171}{MnLargeSymbols}{'171}
\makeatother

\usepackage{eso-pic}

\begin{document}

\title{Beyond Hagedorn: A Harmonic Approach to $T\bar{T}$-deformation}

\author{Jie Gu$^{a}$}

\author{Jue Hou$^{a,c}$}
\thanks{Corresponding author: jue.1.hou@kcl.ac.uk}

\author{Yunfeng Jiang$^{a,b}$}
\thanks{Corresponding author: jinagyf2008@seu.edu.cn}

\affiliation{$^a$School of Physics \& Shing-Tung Yau Center, Southeast University, Nanjing 211189, P. R. China}
\affiliation{$^b$Peng Huanwu Center for Fundamental Theory, Hefei, Anhui 230026, China}
\affiliation{$^c$Department of Mathematics, King’s College London, The Strand, London WC2R 2LS, UK}

\begin{abstract}
We apply harmonic analysis to study the $T\bar{T}$-deformed torus partition function. We first express the CFT partition functions in terms of Maass waveforms, including the Eisenstein series and cusp forms. These basis functions turn out to deform in a very simple way under the $T\bar{T}$-deformation. The spectral decomposition provides a numerically stable and efficient method to compute the partition function at finite values of the deformation parameter $\lambda$, allowing us to clearly resolve the analytic structure of the partition function as a function of $\lambda$. The resulting deformed partition function exhibits a Hagedorn singularity. Building on the harmonic analysis approach, we propose a natural analytic continuation beyond the Hagedorn singularity, which enables us to compute the full partition function for any value of $\lambda$.

\end{abstract}

\maketitle

\noindent{\bf Introduction.} The thermal partition function is a fundamental quantity in the study of quantum field theory (QFT). In 1+1 dimensions, the partition function for a QFT in finite volume has the topology of a torus, one can therefore define the torus partition function \footnote{The partition function depends on both $\tau$ and its complex conjugate $\bar{\tau}$. Following the mathematics literature, we write $Z(\tau)$ instead of $Z(\tau,\bar{\tau})$. This convention applies to all modular functions throughout the paper.}
\begin{align}
Z(\tau)=\sum_n e^{-2\pi R\tau_2 E_n+2\pi \mathrm{i} R\tau_1 P_n},\quad \tau=\tau_1+\mathrm{i}\tau_2,
\end{align}
where the sum runs over all states in the Hilbert space, $E_n$ and $P_n$ denote the energy and momentum of a given state. When the theory is at a critical point, it becomes a conformal field theory (CFT), and the corresponding partition function enjoys a particularly nice property known as \emph{modular invariance}: it is invariant under modular transformations $Z(\gamma\cdot\tau)=Z(\tau)$, where $\gamma\cdot\tau=(a\tau+b)/(c\tau+d)$ with $a,b,c,d\in\mathbb{Z}$ and $ad-bc=1$. This property is remarkably constraining and underlies many beautiful results of two-dimensional CFTs \cite{Ginsparg:1988ui,DiFrancesco:1997nk}.\par

Modular invariance is especially powerful when combined with solvability, which provides analytic control over the torus partition function. Apart from CFTs, another class of solvable but non-local theories in 1+1 dimensions has been intensely studied in recent years. These are the so-called $T\bar{T}$-deformed QFTs \cite{Smirnov:2016lqw,Cavaglia:2016oda}, which preserve an infinite number of symmetries. Taking a CFT as the seed theory, the resulting deformed theory remains solvable, and its spectrum can be written in a simple closed form. More explicitly, for a state with energy and momentum $(E_n,P_n)$, the deformed energy and momentum are $(\mathcal{E}_n(\lambda),P_n)$ with \cite{Smirnov:2016lqw,Cavaglia:2016oda}
\begin{align}
\mathcal{E}_n(\lambda)=\frac{1}{\lambda\pi R}\left(\sqrt{1+2\lambda\pi R E_n+\lambda^2\pi^2 R^2 P_n^2}-1\right).
\end{align}
The solvability of $T\bar{T}$-deformation makes it a valuable example of non-local quantum field theories and a toy model of quantum gravity which can be studied analytically (see \cite{Jiang:2019epa,He:2025ppz} for reviews).

A key quantity of interest in the $T\bar{T}$-deformed theory is the deformed torus partition function, defined by
\begin{align}
\mathcal{Z}(\tau|\lambda)=\sum_n e^{-2\pi R\tau_2\mathcal{E}_n(\lambda)+2\pi\mathrm{i}R\tau_1 P_n}.
\end{align}
It has been shown that the deformed torus partition function is also modular invariant: $\mathcal{Z}(\gamma\cdot\tau|\gamma\cdot\lambda)=\mathcal{Z}(\tau|\lambda)$ with $\gamma\cdot\lambda=\lambda/|c\tau+d|^2$. In fact, it can be shown that among all deformations of the form $(E_n,P_n)\to (\mathcal{H}(E_n,P_n,\lambda),P_n)$ with $\mathcal{H}(E,P,\lambda)$ being an arbitrary function of $E,P$ and $\lambda$, the $T\bar{T}$-deformation is the \emph{unique} one that yields a modular invariant deformed partition function \cite{Aharony:2018bad}. Exploiting modular invariance, one can derive the asymptotic density of states, which is found to interpolate between a Cardy-like behavior $\rho(E)\sim e^{\sqrt{E}}$ for small $\lambda$ and a Hagedorn-like behavior $\rho(E)\sim e^{E}$ for large $\lambda$. Modular invariance has also played a central role in developing an efficient method for computing the series expansion of $\mathcal{Z}(\tau|\lambda)$ as an infinite series in $\lambda$. For fixed $\tau$, this series is asymptotic, and using resurgence theory, the non-perturbative contributions have been identified \cite{Gu:2024ogh,Gu:2025tpy}. Modular invariance has further played an important role in establishing holographic duals of the $T\bar{T}$-deformed CFT \cite{McGough:2016lol,Chen:2019mis,Apolo:2023vnm,Guica:2019nzm,Hirano:2020nwq,Apolo:2023aho,Dei:2024sct}.

Despite substantial progress in understanding the $T\bar{T}$-deformed torus partition function, two outstanding questions remain. The first concerns the analytic properties of $\mathcal{Z}(\tau|\lambda)$ as a function of $\lambda$. Resurgence analysis indicates the existence of a branch cut, but are there other singularities or more intricate structure? One possible approach is to devise a method to evaluate the partition function for finite values of $\lambda$, even if only numerically. This could also be important for other purposes, such as computing the spectral form factor and other correlators of partition functions, which play a crucial role in the study of quantum chaos \cite{Cotler:2016fpe,Dyer:2016pou}. The second issue is that the partition function for $\lambda>0$ is known to develop a Hagedorn singularity at finite $\lambda$ \cite{Datta:2018thy}. Beyond this point, the partition function diverges. Can one find an analytic continuation beyond this singularity? In this work, we propose that \emph{harmonic analysis} provides a powerful tool to address both questions.

By harmonic analysis, we mean expanding the torus partition function in terms of eigenfunctions of the Laplacian $\Delta_{\tau}$ on the Poincaré upper half plane. These eigenfunctions are known as Maass waveforms. The application of Maass forms to CFT has been explored in several works \cite{Benjamin:2021ygh,Benjamin:2022pnx,DHoker:2022dxx,Haehl:2023tkr,Benjamin:2025kvm,DiUbaldo:2023qli,Boruch:2025ilr,Perlmutter:2025ngj}. Their connections with $T\bar{T}$-deformed CFTs have also been discussed previously in \cite{Benjamin:2023nts,Cardy:2022mhn,Godet:2024ich}. 

A key result in the harmonic analysis of modular invariant functions is the Roelcke–Selberg spectral decomposition theorem\cite{rankin1939contributions,selberg1940bemerkungen,zagier1981rankin}, which states that any square integrable modular invariant function $f(\tau)\in L^2(\mathcal{F})$ can be expanded in terms of real-analytic Eisenstein series $E_s(\tau)$ with $\text{Re}(s)=1/2$ and the Maass cusp forms $\nu_n(\tau)$. However, an immediate difficulty in applying this theorem to CFTs is that the partition functions $Z(\tau)$ are \emph{not} square integrable on the fundamental domain. A common remedy is to multiply the partition function by appropriate powers of $\sqrt{\tau_2}|\eta(\tau)|^2$ ($\eta(\tau)$ being the Dedekind $\eta$-function) to ``regularize'' it, and then apply the theorem to the resulting quantity, known as the primary partition function. For $T\bar{T}$-deformed theories, however, the primary partition function is not a convenient quantity to work with because it does not satisfy a simple flow equation. 

In this work, we adopt a different strategy which separates the square-integrable part (denoted by $Z_R$) from the non-square-integrable part (denoted by $Z_E$) in a modular invariant way. Both parts can be expanded in terms of Eisenstein series and cusp forms, which are deformed in a simple way under $T\bar{T}$-deformation. The Hagedorn singularity of the deformed partition function originates from $Z_E$, and we propose a natural analytic continuation beyond the Hagedorn point. In this way, we develop a numerically stable and efficient way to compute the deformed partition function at finite values of $\lambda$, which uncovers the analytic properties of the deformed partition function.

\vspace{0.5cm}

\noindent{\bf Spectral expansion of CFT.}  To perform the harmonic analysis of a modular invariant function $f(\gamma\cdot\tau)=f(\tau)$, we expand it in terms of eigenfunctions of the Laplacian on the Poincar\'e upper half plane $\Delta_{\tau}=-\tau_2^2(\partial_{\tau_1}^2+\partial_{\tau_2}^2)$ ($\tau_1\in(-\infty,\infty)$, $\tau_2\ge 0$). 

\vspace{0.3cm}
\noindent{\it - Maass waveforms.}
The spectrum of $\Delta_{\tau}$ contains both continuous and discrete parts. The discrete part is called the Maass cusp form $\nu_n(\tau)$, ($n=0,1,\ldots$) with $\nu_0=\sqrt{3/\pi}$. The cusp forms decay exponentially around the cusp,
\begin{align}
\nu_n(\tau)\sim e^{-2\pi \tau_2},\qquad \tau_2\to\infty\,.
\end{align}
The continuous part of the spectrum is spanned by real-analytic Eisenstein series $E_s(\tau)$ where $s$ is a complex number. The Eisenstein series has the following asymptotic behavior near the cusp
\begin{align}
\label{eq:asymEs}
E_s(\tau)\sim \tau_2^s+\frac{\Lambda(1-s)}{\Lambda(s)}\tau_2^{1-s},\qquad \tau_2\to\infty.
\end{align}
Explicit expressions and main properties of these functions can be found in SM.

If $f(\tau)$ is square integrable on the fundamental domain $\mathcal{F}=\mathbb{H}/\Gamma$ where $\mathbb{H}$ is the upper half-plane and $\Gamma=SL(2,\mathbb{Z})$, it can be decomposed as follows
\begin{align}
f(\tau)=\sum_{n=0}^{\infty}\frac{(f,\nu_n)}{(\nu_n,\nu_n)}\nu_n(\tau)+\frac{1}{4\pi\mathrm{i}}\int_{\text{Re}\,s=\frac{1}{2}}(f,E_s)E_s(\tau)\mathrm{d}s
\end{align}
where $(f,g)$ is the modular invariant Petersson inner product defined by
\begin{align}
(f,g)\equiv\int_{\mathcal{F}}\frac{\mathrm{d}\tau_1\mathrm{d}\tau_2}{\tau_2^2}f(\tau)\overline{g(\tau)}\,.
\end{align}
This result is known as the Roelcke-Selberg (RS) spectral decomposition theorem (see \emph{e.g.}, \cite{terras2013harmonic}). 

\vspace{0.3cm}
\noindent{\it - Modular invariant regularization.} For concreteness we will consider rational CFTs such as free theories and minimal models. The partition function $Z(\tau)$ is not square integrable. Near the cusp, which corresponds to the low temperature limit, it is dominated by the vacuum and $Z(\tau)\sim e^{-2\pi \tau_2 RE_0}$, which is growing exponentially if $E_0<0$. For unitary CFTs, we have $E_0R=-\mathrm{c}/12<0$ with $\mathrm{c}$ being the central charge. For a non-unitary CFT such as the Lee-Yang model, we have $\mathrm{c}=-22/5<0$, but the ground state energy $E_0 R=h_{\text{min}}+\bar{h}_{\text{min}}-\mathrm{c}/12=-1/30$ is again negative, leading to exponential growth for $\tau_2\gg 1$. In what follows, we shall denote for both cases the ground energy by $E_0R=-\tc/12$ where $\tc>0$. Let us define the following series
\begin{align}
\label{eq:defZE}
Z_E(\tau,\tc)=1-\frac{2\pi \tc}{12}\kappa+\frac{2\pi \tc}{12}E_1^r(\tau)+\sum_{k=2}^{\infty}\frac{1}{k!}\left(\frac{2\pi \tc}{12}\right)^k E_k(\tau)\,.
\end{align}

Note that around $s=1$ the Eisenstein series is singular
\begin{align}
E_{1+\epsilon}(\tau)=\frac{3}{\pi\epsilon}+\tau_2-\frac{3}{\pi}\log\tau_2+\kappa+\mathcal{O}(\epsilon)
\end{align}
with $\kappa=\frac{6}{\pi}\left(\gamma-\log 2-\frac{6}{\pi^2} \zeta^{\prime}(2)\right)\approx 0.867132$, where $\gamma$ is Euler’s constant and $\zeta(z)$ is
Riemann zeta function. We thus define the regularized quantity $E_1^r(\tau)\equiv\lim_{\epsilon\to 0}(E_{1+\epsilon}-\frac{3}{\pi\epsilon})$ in \eqref{eq:defZE}. Using the asymptotic behavior of the Eisenstein series \eqref{eq:asymEs}, we find that $Z_E(\tau,\tc)\sim e^{2\pi \tc\tau_2/12}$ near the cusp, which is the same as $Z(\tau)$. The key point is that the regularized partition function
\begin{align}
Z_R(\tau)=Z(\tau)-Z_E(\tau,\tc)
\end{align}
is modular invariant and square integrable, which can be decomposed using the RS theorem. If there are more states with negative energy (including degeneracies of the ground state), we can subtract them by proper infinite series of Eisenstein series in the same way. Denoting $\mathsf{v}_n\equiv(Z_R,\nu_n)/(\nu_n,\nu_n)$ and $\mathsf{e}_s\equiv(Z_R,E_s)$, we arrive at the following spectral decomposition for the torus partition function of a CFT
\begin{equation}
\label{eq:RSdecomZ}
\begin{aligned}
Z(\tau)=&1-\frac{2\pi \tc}{12}\kappa+\frac{2\pi \tc}{12}E_1^r(\tau)+\sum_{k=2}^{\infty}\frac{1}{k!}\left(\frac{2\pi \tc}{12}\right)^kE_k(\tau)\\
&+\sum_{n=0}^{\infty}\mathsf{v}_n\,\nu_n(\tau)+\frac{1}{4\pi\mathrm{i}}\int_{\text{Re}\,s=\frac{1}{2}}\mathsf{e}_s\,E_s(\tau)\mathrm{d}s\,.
\end{aligned}
\end{equation}
Here $\mathsf{v}_n$ and $\mathsf{e}_s$ are numbers and can be computed numerically for a given torus partition function of a CFT.
\vspace{0.5cm}

\noindent{\bf $T\bar{T}$-deformation.} 
Now we consider the $T\bar{T}$-deformed torus partition function.

\vspace{0.3cm}
\noindent{\it - An integral transform.}
The deformed and the undeformed partition functions are related by the following integral transform \cite{Dubovsky:2018bmo,Hashimoto:2019wct}
\begin{align}
\label{eq:intTransform}
\mathcal{Z}(\tau|\lambda)=\frac{\tau_2}{\pi\lambda}\int_{\mathbb{H}}\frac{\mathrm{d}^2\zeta}{\zeta_2^2}e^{-\frac{|\zeta-\tau|^2}{\lambda\zeta_2}}Z(\zeta)\,.
\end{align}
Motivated by this, we define the following integral transform for any modular function $f(\tau)$
\begin{align}
\label{eq:defTchi1}
\mathcal{T}_{\chi}[f]\equiv\frac{\chi}{\pi}\int_{\mathbb{H}}\frac{\mathrm{d}^2\zeta}{\zeta_2^2}\,e^{-\chi\frac{|\tau-\zeta|^2}{\tau_2\zeta_2}}f(\zeta)\,,
\end{align}
where $\chi=\tau_2/\lambda$ is invariant under modular transformation. The $\mathcal{T}_{\chi}$-transform has the following property
\begin{itemize}
\item It is linear. For two modular functions $f_1(\tau)$ and $f_2(\tau)$, we have
\begin{align}
\mathcal{T}_{\chi}[\alpha f_1+\beta f_2]=\alpha\,\mathcal{T}_{\chi}[f_1]+\beta\,\mathcal{T}_{\chi}[f_2]\,.
\end{align}
\item It commutes with modular transformation, \emph{i.e.}
\begin{align}
\label{eq:defTchi}
\mathcal{T}_{\chi}[\gamma\cdot f]=\gamma\cdot\mathcal{T}_{\chi}[f]\,.
\end{align}
\end{itemize}
The first property is obvious and the proof of the second property will be given in SM. 

\vspace{0.3cm}
\noindent{\it - $\mathcal{T}_{\chi}$-transform of Maass forms.}
Using the property of the $\mathcal{T}_{\chi}$-transform, we have the following result: The $\mathcal{T}_{\chi}$-transform of $E_s(\tau)$ and $\nu_n(\tau)$ are multiplicative
\begin{align}
\label{eq:actionTMaass}
\mathcal{T}_{\chi}[E_s]=C_s(\chi)\,E_s(\tau),\quad \mathcal{T}_{\chi}[\nu_n]=C_{s_n}(\chi)\nu_n(\tau)
\end{align}
with $C_j(\chi)=\frac{2}{\sqrt{\pi}}e^{2\chi}\sqrt{\chi}K_{j-1/2}(2\chi)$\,. As a corollary, its action on a constant is invariant $\mathcal{T}_{\chi}[c]=c$ and its action on $E_1^r(\tau)$ is given by
\begin{align}
\label{eq:TchiE1}
\mathcal{T}_{\chi}[E_1^r]=E_1^r(\tau)+\frac{3}{\pi}U(1,1,4\chi)
\end{align}
where $U(a,b,z)$ is the confluent hypergeometric function. The proof of \eqref{eq:actionTMaass} is given in SM.

\vspace{0.3cm}
\noindent{\it - Deformed partition function.}
The deformed partition function is given by
\begin{align}
\label{eq:mainHarm}
\mathcal{Z}(\tau|\lambda)=&\,\mathcal{T}_{\chi}[Z]=\mathcal{T}_{\chi}[Z_E]+\mathcal{T}_{\chi}[Z_R]\\\nonumber
=&\,1-\frac{2\pi \tc}{12}\kappa+\frac{2\pi \tc}{12}\left(E_1^r(\tau)+\frac{3}{\pi}U(1,1,4\chi)\right)\\\nonumber
&\,+\sum_{k=2}^{\infty}\frac{1}{k!}\left(\frac{2\pi \tc}{12}\right)^k C_k(\chi)E_k(\tau)+\sum_{n=1}^{\infty}\mathsf{v}_nC_{s_n}(\chi)\nu_n(\tau)\\\nonumber
&\,+\nu_0+\frac{1}{4\pi\mathrm{i}}\int_{\text{Re}\,s=\frac{1}{2}}\mathsf{e}_s C_s(\chi)E_s(\tau)\mathrm{d}s.
\end{align}
As can be seen, all dependence on the deformation parameter $\lambda$ is encapsulated in the functions $C_j(\chi)$ and $U(1,1,4\chi)$. This is one of the main results of the present work. The above formula greatly facilitates the study of the analytic dependence on $\lambda$. 

The expansion \eqref{eq:mainHarm} provides a useful representation for the numerical evaluation of the deformed partition function at finite $\tau$ and $\lambda$. The numerical coefficients $\mathsf{v}_n$ and $\mathsf{e}_s$ are first evaluated for a sufficiently large set of $n$ and $s$, which are computed using the seed CFT and is independent of the $T\bar{T}$-deformation. Interestingly, it turns out for all the models under consideration we have $\mathsf{v}_n=0$. The numerical evaluation of the Maass forms and $C_j(\chi)$ can be carried out straightforwardly. By discretizing the integral in \eqref{eq:mainHarm} as a sum with an appropriate truncation, we obtain numerical results for the deformed partition function with high accuracy. In principle, one could also use \eqref{eq:intTransform} directly to evaluate the deformed partition function. However, the exponential kernel is numerically unstable which makes it difficult to use for obtaining reliable results. The representation \eqref{eq:mainHarm} offers a numerically stable and efficient alternative. Details of the numerical implementation will be presented elsewhere \cite{ToAppear}. We have explicitly performed the calculation for several concrete examples, including the free compact boson, the free fermion, and the Lee-Yang model.

We cross-check our results against the optimal truncation (OT) of the infinite series expansion of $\mathcal{Z}(\tau|\lambda)$. Although the series $\mathcal{Z}(\tau|\lambda)=\sum_{n=0}^{\infty}Z_n\lambda^n$ is asymptotic, it can be truncated at a finite order to obtain an approximation of the deformed partition function. There exists an optimal order $n=n^*$ for this purpose, while adding more terms beyond this order actually worsens the approximation. The error of the optimal truncation can be estimated. Within the estimated errors, the results from harmonic analysis (HA) and optimal truncation are in good agreement, as shown in Tables~\ref{tab:Fermion-HO-OT} and~\ref{tab:LY-HO-OT}.\begin{table}[h!]
    \centering
    \begin{tabular}{ccc}
    \hline\hline
        $\tau | \lambda$ & \text{HA} & \text{OT} \\ \hline
        $\ri | 0.3$ & $3.84917+\mathcal{O}(10^{-6})$ & $3.848914+\mathcal{O}(10^{-3})$\\
       $ \ri | 0.1$ & $3.81082+\mathcal{O}(10^{-6})$ & $3.810821+\mathcal{O}(10^{-7})$\\
        $0.23+1.8\ri | 0.3$ & $4.04215+\mathcal{O}(10^{-6})$ & $4.042149+\mathcal{O}(10^{-6})$\\
       $ 0.23+1.8\ri | 0.1$ & $4.00288+\mathcal{O}(10^{-6}) $& $4.002876+\mathcal{O}(10^{-7})$\\
         \hline\hline
    \end{tabular}
    \caption{Comparison of results of harmonic analysis (HA) and optimal truncation (OT) of perturbative series for the deformed partition function of the free fermion model. The estimated error margins are also indicated by the $\mathcal{O}$ notation.}
    \label{tab:Fermion-HO-OT}
\end{table}

\begin{table}[h!]
    \centering
    \begin{tabular}{ccc}
    \hline\hline
        $\tau|\lambda$ & \text{HA} & \text{OT} \\\hline
        $\ri | 0.3$ & $1.37673+\mathcal{O}(10^{-6})$ & $1.376181+\mathcal{O}(10^{-3})$\\
       $ \ri | 0.1$ & $1.35133+\mathcal{O}(10^{-6})$ & $1.351327+\mathcal{O}(10^{-7})$\\
        $0.23+1.8\ri | 0.3$ & $1.49402+\mathcal{O}(10^{-6})$ & $1.494015+\mathcal{O}(10^{-6})$\\
        $0.23+1.8\ri | 0.1$ & $1.48041+\mathcal{O}(10^{-6})$ & $1.480407+\mathcal{O}(10^{-7})$\\
         \hline\hline
    \end{tabular}
    \caption{Comparison of results of harmonic analysis (HA) and optimal truncation (OT) of perturbative series for the $T\bar{T}$-deformed partition function of the Lee-Yang model. The estimated error margins are also indicated by the $\mathcal{O}$ notation.}
    \label{tab:LY-HO-OT}
\end{table}

\begin{figure}
    \centering
    \subfloat[Real part]{\includegraphics[width=0.5\linewidth]{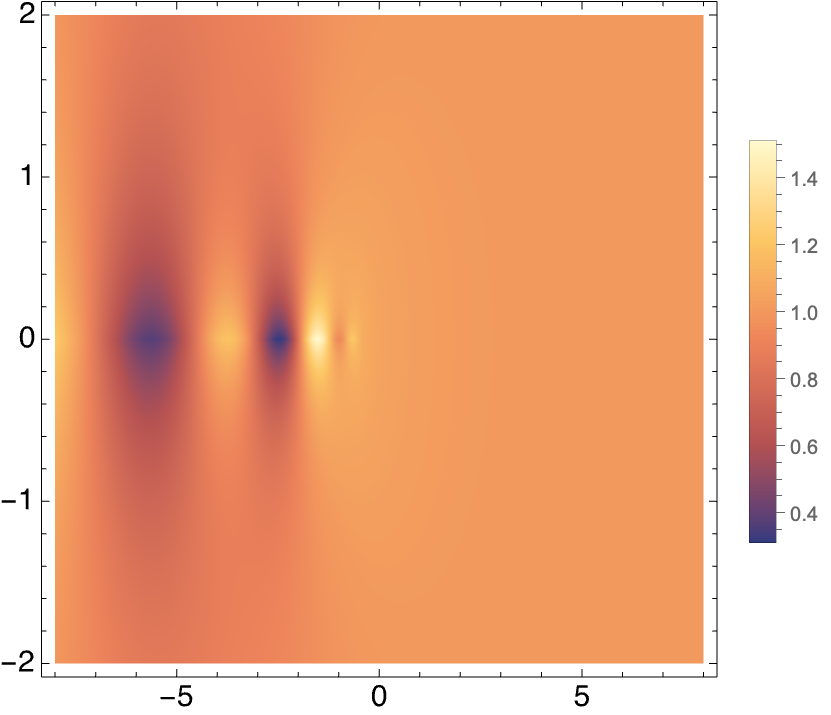}}
    \subfloat[Imaginary part]{\includegraphics[width=0.5\linewidth]{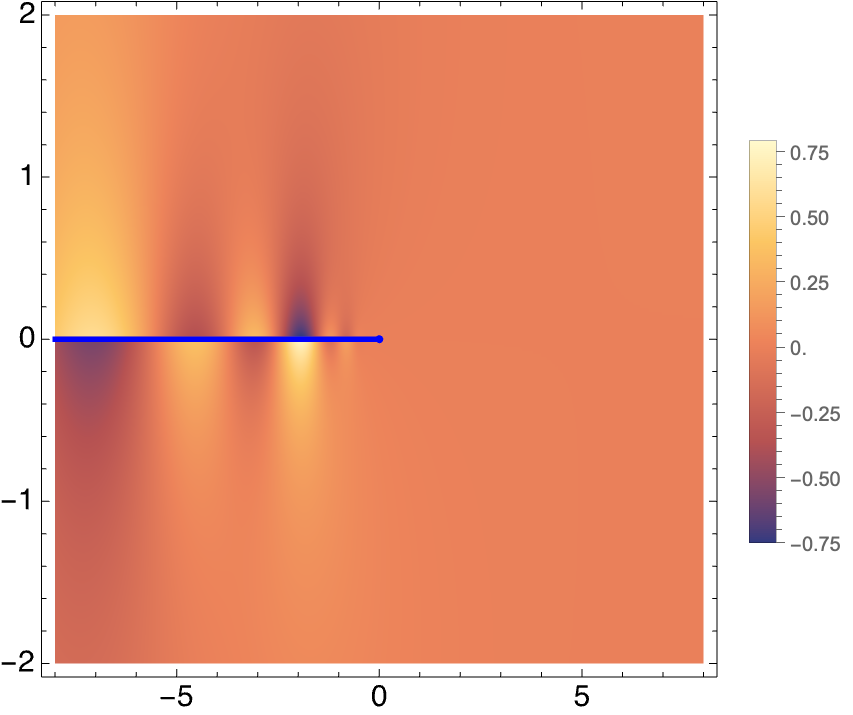}}
    \caption{Density plots of the real part (a) and the imaginary part (b) of $\mathcal{T}_\chi[Z_R](\tau)$ for free fermion for $\tau = 0.12+\ri 1.15$ and $\lambda$ in the range from $-8-\ri 2$ (bottom left) to $7+\ri 2$ (top right). The plot is brighter for higher value and darker for lower value. The branch cut along the negative real axis is clearly visible in the imaginary plot and is indicated by the blue line.}
    \label{fig:FF-denplot}
\end{figure}

\begin{figure}
    \centering
    \subfloat[Real part]{\includegraphics[width=0.5\linewidth]{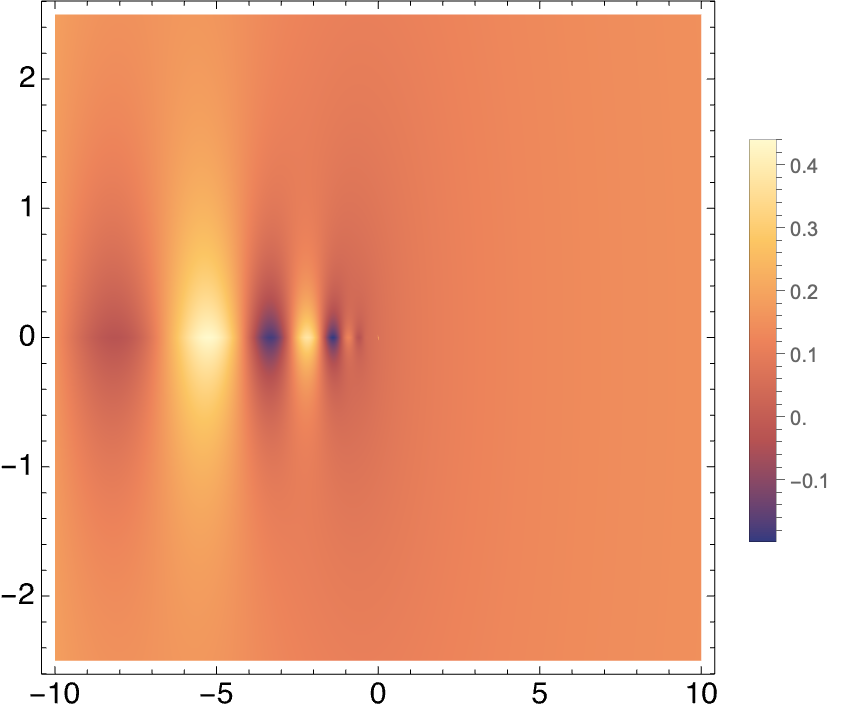}}
    \subfloat[Imaginary part]{\includegraphics[width=0.5\linewidth]{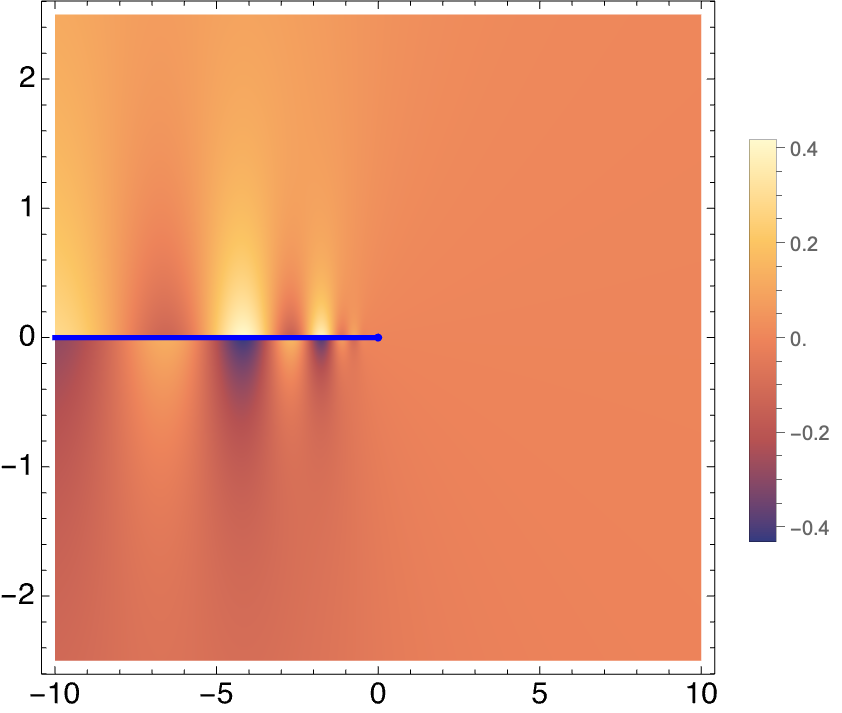}}
    \caption{Density plots of the real part (a) and the imaginary part (b) of $\mathcal{T}_\chi[Z_R](\tau)$ for Lee-Yang for $\tau = 0.12+\ri 1.15$ and $\lambda$ in the range from $-10-\ri 2.5$ (bottom left) to $10+\ri 2.5$ (top right). The plot is brighter for higher value and darker for lower value. The branch cut along the negative real axis is clearly visible in the imaginary plot and is indicated by the blue line.}
    \label{fig:LY-denplot}
\end{figure}

Using the HA method, we can compute the deformed partition function at finite values of $\tau$ and $\lambda$. We find that $\mathcal{T}_{\chi}[Z_R]$ exhibits no divergences and can be computed for any $\tau$ and $\lambda$, as illustrated in Fig.~\ref{fig:FF-denplot} and \ref{fig:LY-denplot} respectively for the Ising model and Lee-Yang model. The functions $C_j(\chi)=C_j(\tau_2/\lambda)$ possess a branch cut along $\lambda\in(-\infty,0]$, which is clearly visible in the plot of the deformed partition function.

In contrast, $\mathcal{T}_{\chi}[Z_E]$ is convergent only within a certain range of $\chi$. This is the Hagedorn singularity of the deformed partition function. It originates from the Hagedorn-like asymptotic growth of the high-energy density of states, which, under the $S$-modular transform, is mapped to the vacuum sector and is associated with the exponentially growing part of the undeformed CFT partition function. It is therefore expected to appear in $\mathcal{T}_{\chi}[Z_E]$, which is essentially the $T\bar{T}$-deformation of that exponentially growing part.

\begin{figure}
    \begin{center}
    \subfloat[Real part]{\includegraphics[width=0.5\linewidth]{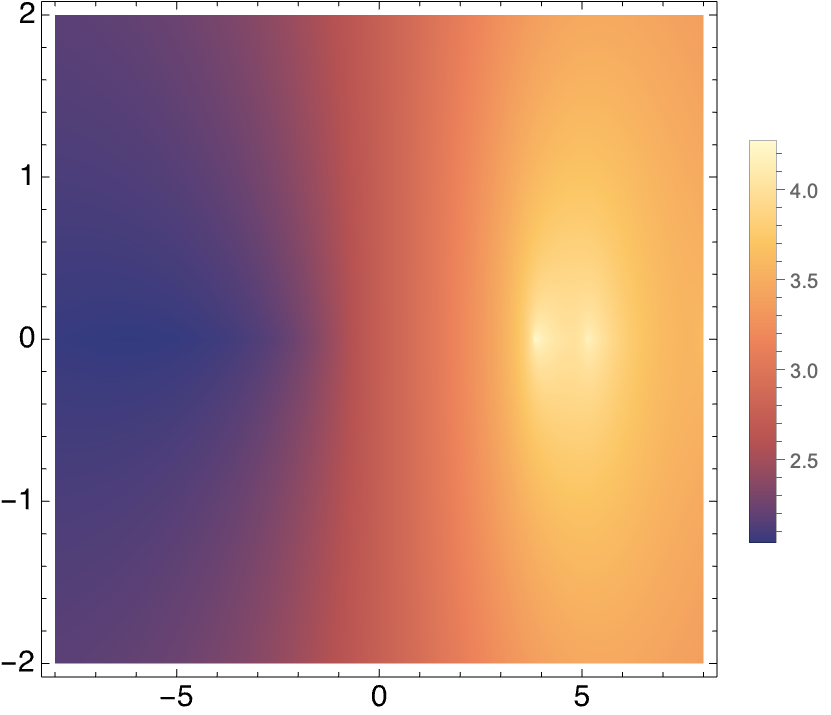}}
    \subfloat[Imaginary part]{\includegraphics[width=0.5\linewidth]{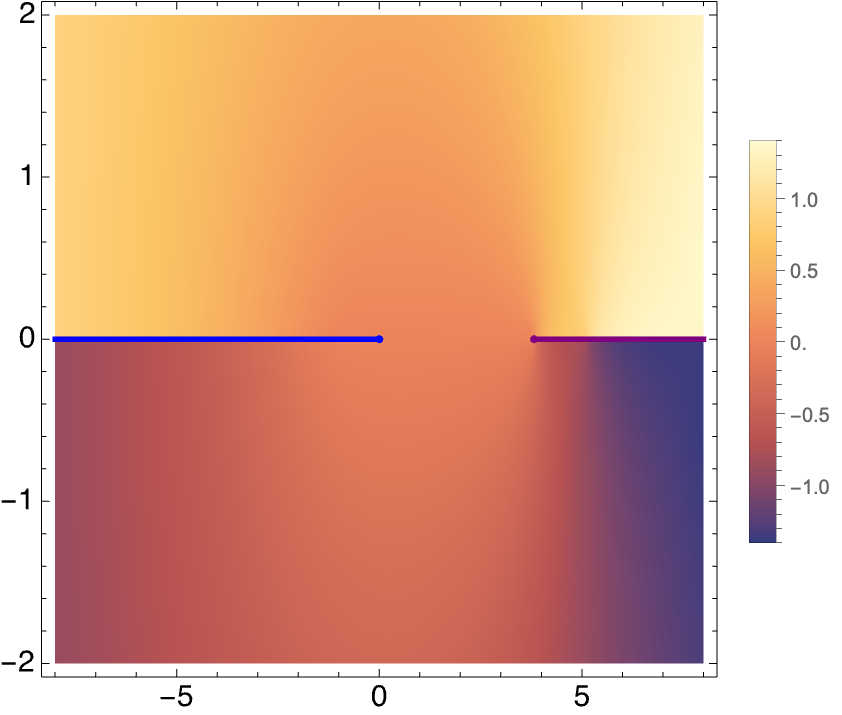}}\\
    \subfloat[Just below the real axis]{\includegraphics[width=0.5\linewidth]{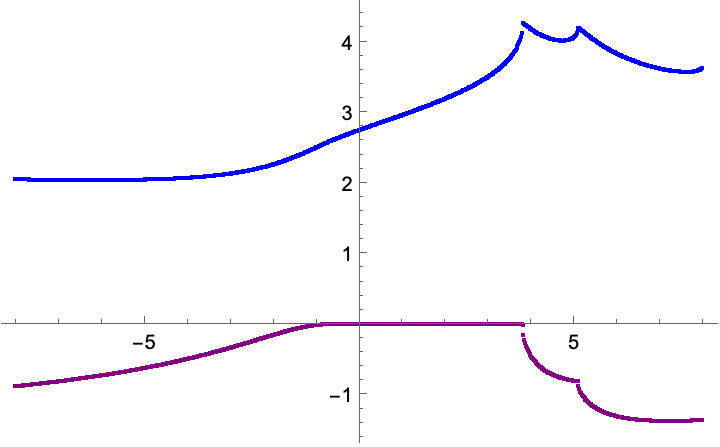}}
    \end{center}
    \caption{Density plots of the real part (a) and the imaginary part (b) as well as the plot of both the real (blue) and the imaginary (purple) parts of $\widetilde{\mathcal{T}_\chi}[Z_E](\tau)$ just below the real axis (c) for free fermion with $\tau = 0.12+\ri 1.15$ and $\lambda$ in the range from $-8-\ri 2$ (bottom left) to $8+\ri 2$ (top right). In the density plots, the plot is brighter for higher value and darker for lower value. In the imaginary density plot (b), the branch cut along the negative real axis due to the hypergeometric function $U(1,1,4\chi)$ and the one on the positive real axis from $6/(\pi \bar{c})\approx 3.82$ due to Hagedorn singularity are clearly visible and are indicated by the blue and purple lines. 
    In plot (c), a second branch point on the positive real axis at $6|\tau|^2/(\pi\bar{c})\approx 5.11$ can also been seen.
    }
    \label{fig:FF-Ze-denplot}
\end{figure}

\vspace{0.5cm}

\noindent{\bf Hagedorn and beyond.} 
We now turn to the Hagedorn singularity of the deformed partition function.

\vspace{0.3cm}
\noindent{\it - Asymptotic behavior.} To better understand the nature of the singularity, consider the behavior of $\mathcal{T}_{\chi}[Z_E]$ near the cusp $\tau_2\gg 1$. Using the asymptotics of the Eisenstein series, we obtain
\begin{align}
\mathcal{T}_{\chi}[Z_E]=&\,\frac{\bar{c}}{2}(U(1,1,4\chi)-\log\tau_2)\\\nonumber
&\,+\sum_{k=0}^{\infty}C_k(\chi)\frac{1}{k!}\left(\frac{2\pi \tc\tau_2}{12}\right)^k+\cdots
\end{align}
where the ellipsis denotes subleading terms in $\tau_2$. Focusing on the infinite sum
\begin{align}
&\sum_{k=0}^{\infty}C_k(\chi)\frac{1}{k!}\left(\frac{2\pi \tc\tau_2}{12}\right)^k\\\nonumber
&=\frac{2}{\sqrt{\pi}}e^{2\chi}\sqrt{\chi}\sum_{k=0}^{\infty}\frac{1}{k!}\left(\frac{2\pi \tc\lambda}{12}\chi\right)^kK_{k-\frac{1}{2}}(2\chi)\,,
\end{align}
the right-hand side can be evaluated using the multiplication theorem for Bessel functions (see \emph{e.g.} \cite{gradshteyn2014table})
\begin{align}
\xi^{-\nu}K_{\nu}(\xi z)=\sum_{k=0}^{\infty}\frac{1}{k!}\left(\frac{(1-\xi^2)z}{2}\right)^k K_{\nu+k}(z)\,,
\end{align}
which converges for $|1-\xi^2|<1$. Setting $1-\xi^2=2\pi \tc\lambda/12$ and $z=2\chi$, we obtain
\begin{align}
\label{eq:productForC}
\sum_{k=0}^{\infty}C_k(\chi)\frac{1}{k!}\left(\frac{2\pi \tc\tau_2}{12}\right)^k=\exp\left(2\chi(1-\xi)\right)
\end{align}
within the convergence region $|\pi \tc\lambda/6|<1$. For $|\pi \tc \lambda/6|\ge 1$, the series \eqref{eq:productForC} ceases to converge at the boundary, for positive real $\lambda$ this gives the Hagedorn divergence. The right-hand side remains well-defined even for $\lambda$ outside the convergence region, and we can take it as analytic continuation of the left-hand side.

\vspace{0.3cm}
\noindent{\it - Analytic continuation.}  The above analysis provides a natural avenue for analytically continuing the partition function beyond the Hagedorn singularity. Since $\mathcal{T}_{\chi}[Z_R]$ is regular, we focus on $\mathcal{T}_{\chi}[Z_E]$. Using the fact that the Eisenstein series can be expressed as a Poincar\'e series,
\begin{align}
E_s(\tau)=\frac{1}{2}\sum_{\gamma\in \Gamma/\Gamma_{\infty}}(\text{Im}\,\gamma\cdot\tau)^s=\frac{1}{2}\sum_{\text{GCD}(c,d)=1}\frac{\tau_2^s}{|c\tau+d|^{2s}}
\end{align}
where $\Gamma_{\infty}$ is generated by $\left(\begin{array}{ll}1 & 1 \\ 0 & 1\end{array}\right)$,
we can write
\begin{align}
\mathcal{T}_{\chi}[Z_E]=&\,\sum_{k=2}^{\infty}\frac{1}{k!}\left(\frac{2\pi \tc}{12}\right)^kC_k(\chi) E_k(\tau)+\cdots\\\nonumber
=&\,\frac{1}{2}\sum_{k=2}^{\infty}\frac{1}{k!}\left(\frac{2\pi \tc}{12}\right)^kC_k(\chi) \sum_{\gamma\in\Gamma/\Gamma_{\infty}}(\text{Im}\,\gamma\cdot\tau)^k+\cdots .
\end{align}
Now we exchange the order of the two summations. If $\mathcal{T}_{\chi}[Z_E]$ is absolutely convergent, this interchange is justified; otherwise, we adopt it as \emph{a prescription for analytic continuation}, yielding
\begin{align}
\widetilde{\mathcal{T}_{\chi}}[Z_E]&=\,1-\frac{2\pi \tc}{12}\kappa+\frac{2\pi \tc}{12}\left(E_1^r(\tau)+\frac{3}{\pi}U(1,1,4\chi) \right)\nonumber\\
&\,+\frac{1}{2}\sum_{\gamma\in\Gamma/\Gamma_{\infty}}\left[\exp\left(2\chi\left(1-\sqrt{1-\frac{2\pi \tc}{12\chi}\text{Im}(\gamma\cdot\tau)} \right)\right)\right.\nonumber\\
&\qquad\qquad\left. -1-\frac{2\pi \tc\,\text{Im}(\gamma\cdot\tau)}{12}\right]
\label{eq:TZE}
\end{align}
where we have used \eqref{eq:productForC}. The Poincar\'e sum in the second line converges, whose proof can be found in the SM, and we take $\widetilde{\mathcal{T}_{\chi}} [Z_E]$ as the analytic continuation of ${\mathcal{T}_{\chi}}[Z_E]$. Consequently, the analytically continued partition function is given by
\begin{align}
\widetilde{\mathcal{Z}}(\tau|\lambda)=\widetilde{\mathcal{T}_{\chi}}[Z_E]+\mathcal{T}_{\chi}[Z_R]\,.
\end{align}
This constitutes another main result of the present work. We have computed the Poincar\'e sum numerically and confirmed that it indeed converges rapidly.

\vspace{0.3cm}
\noindent{\it - Hagedorn singularities.}
After analytic continuation, the partition function is no longer divergent. The Hagedorn singularities are now manifest as branch points of $\widetilde{\mathcal{T}_{\chi}}[Z_E]$. It has a branch point whenever
\begin{equation}\label{eq:lmb-br}
    \lambda = \lambda_\gamma \equiv \frac{6}{\pi \tc}\frac{{\rm Im}\tau}{{\rm Im}(\gamma\cdot\tau)},\quad \gamma \in \Gamma/\Gamma_\infty,
\end{equation}
and there are infinitely many of them. They are all located on the positive real axis in the complex $\lambda$-plane, and are bounded from below in the range $[\lambda_m,\infty)$, where the minimal branch point $\lambda_m$ is reached when $\gamma$ in \eqref{eq:lmb-br} is such that $\gamma\cdot \tau$ is in the fundamental domain so that the denominator is maximum \footnote{Note that even though there are infinitely many branch points in the range  $[\lambda_m,\infty)$, they are not dense. The location of branch points $\lambda_\gamma$ can be written as 
$\lambda_\gamma = 6/(\pi\tc)((c\tau_1+d)^2+c^2\tau_2^2)$ where $c,d$ are coprime integers, and it is then easy to show that
there are finitely many $\lambda_\gamma$ small than $T$ for any $T > 0$.}. This is illustrated in Fig.~\ref{fig:FF-Ze-denplot} for the Ising model. The form of $Z_E$ and $\widetilde{\mathcal{T}_{\chi}}[Z_E]$ are universal and the only difference between different CFTs is the value of $\bar{c}$. Therefore the analytic structure for a generic CFT is very similar to the one shown in Fig.~\ref{fig:FF-Ze-denplot}.

\vspace{0.5cm}

\noindent{\bf Discussions and outlook.}
In this work, we have demonstrated that harmonic analysis provides a powerful framework for studying the $T\bar{T}$-deformed partition function. A key result is the derivation of a Roelcke–Selberg-type spectral decomposition of the deformed partition function. To achieve this, we first separate the CFT partition function into two modular-invariant parts. One part is not square-integrable and grows exponentially near the cusp, yet it has a relatively simple structure and can be expressed as an infinite series of Eisenstein series. The remaining part is square-integrable and admits a standard Roelcke–Selberg spectral decomposition. The $T\bar{T}$-deformed partition function is obtained via the $\mathcal{T}_{\chi}$-transform of the undeformed CFT partition function. We show that the $\mathcal{T}_{\chi}$-transform acts simply on Maass forms, leading to a clean spectral decomposition of the deformed partition function. This representation enables a numerical approach to compute the partition function at finite $\tau$ and $\lambda$. Moreover, the Hagedorn singularity is identified straightforwardly, and we propose a natural analytic continuation beyond it.

This work opens several avenues for future investigation. A natural generalization is to study quantities closely related to the partition function. For instance, the torus partition function of the single-trace $T\bar{T}$-deformed theory  can be addressed using harmonic analysis. In that case, one must consider the symmetric product orbifold of the $T\bar{T}$-deformed partition function \cite{Hashimoto:2019wct,Apolo:2023aho}. The resulting partition function can be obtained by acting with Hecke operators on the double-trace deformed partition functions studied here. A notable advantage is that Maass forms are eigenfunctions of the Hecke operators, so the harmonic analysis approach provides a natural representation for studying the single-trace deformed partition function and may facilitate the analysis in the large-$N$ limit. Another quantity of significant interest is the spectral form factor, which requires analytic continuation of the partition function. Within our approach, numerical computation of the spectral form factor is straightforward. Given that the spectral form factor is an important diagnostic of quantum chaos, it would be interesting to extend the present method to chaotic CFTs. In principle, once the spectral data $\mathsf{v}_n$ and $\mathsf{e}_s$ are obtained, the remainder of the analysis follows similarly. This would open a pathway to investigate how $T\bar{T}$-deformation affects quantum chaos at finite $\lambda$. Some of these questions are currently under investigation and will be reported elsewhere \cite{ToAppear}.

Another important direction is to explore whether similar approaches can be applied to modular covariant quantities. One example is the partition function of $J\bar{T}$-deformed CFTs \cite{Guica:2017lia}, whose properties have been studied in some depth \cite{Aharony:2018ics,Apolo:2019yfj}. While there are similarities to the $T\bar{T}$ case, there are also important differences. The resulting object transforms in a manner reminiscent of a Jacobi form, and the corresponding integral transform is known. Does there exist a natural basis of functions in terms of which such partition functions can be expanded? More generally, one may ask the same question for the combined $T\bar{T}+J\bar{T}+T\bar{J}$ deformation \cite{Chakraborty:2019mdf,Hashimoto:2019wct,Apolo:2021wcn}. These remain intriguing open problems for future work.

\vspace{0.5cm}

\vspace{0.5cm} 
\noindent{\bf Acknowledgements.} 
We are grateful to Nikolay Gromov for the inspiring discussions on separation of variables, and also to Christian Ferko, Gang Yang, Roberto Tateo, and Jie-qiang Wu for their helpful comments.
We would also like to thank Albert and Monica, our two workstations, for working day and night to produce the data for the three figures.
The work of Y.J. is supported by National Natural Science Foundation of China through Grant No.12575073. This work is also supported by the NSFC Grant No. 12247103. J.G. is supported by National Natural Science Foundation of China through Grant No. 12375062.

\appendix
\section{Maass forms}
We give definition and basic properties of the Maass forms in this appendix.
\paragraph{Eisenstein series.} The real-analytic Eisenstein series $E_s(\tau)$ span the continuous spectrum of the Laplacian and satisfies
\begin{align}
\Delta_{\tau}E_s(\tau)=s(1-s)E_s(\tau)\,.
\end{align}
It can be written as a Poincar\'e series
\begin{align}
E_s(\tau)=\frac{1}{2}\sum_{\text{GCD}(c,d)=1}\frac{\tau_2^s}{|c\tau+d|^{2s}}\,
\end{align}
for $\text{Re}(s)>1$ and by analytic continuation for other values of complex number $s$. It has the following Fourier expansion
\begin{equation}
\begin{aligned}
E_s(\tau)=&\tau_2^s+\frac{\Lambda(1-s)}{\Lambda(s)}\tau_2^{1-s}\\
&+\frac{4\sqrt{\tau_2}}{\Lambda(s)}\sum_{n=1}^{\infty}\frac{\sigma_{1-2s}(n)}{n^{s-\frac{1}{2}}}K_{s-\frac{1}{2}}(2\pi n\tau_2)\cos(2\pi n\tau_1)\,,
\end{aligned}
\end{equation}
where $\Lambda(s)=\pi^{-s}\Gamma(s)\zeta(2s)$ is the symmetrized version of Riemann zeta function and $\sigma_s(n)$ is the divisor function defined by
\begin{align}
\sigma_s(n)=\sum_{0<l|n}l^s\,,
\end{align}
with the sum runs over all positive divisors $l$ of $n$.

\paragraph{Maass cusp forms.} The discrete spectrum of the Laplacian is spanned by the constant mode  $\nu_0=\sqrt{3/\pi}$ and the Maass cusp forms $\nu_n(\tau)$, ($n=1,\ldots$) which satisfy
\begin{align}
\Delta_{\tau}\nu_n(\tau)=s_n(1-s_n)\nu_n(\tau)
\end{align}
where $s_n=\frac{1}{2}+\ri\,t_n$ with $t_n>0$. They satisfy
\begin{align}
\int_{-1/2}^{1/2}\nu_n(\tau)\mathrm{d}\tau_1=0\,,\quad n\ge 1\,.
\end{align}
The cusp forms can be classified by their parity under the transformation $\tau\to-\bar{\tau}$. The even and odd forms, denoted by $\nu_n^+(\tau)$ and $\nu_n^-(\tau)$ respectively, satisfy
\begin{align}
\nu_n^+(\tau)=\nu_n^+(-\bar{\tau}),\quad \nu_n^-(\tau)=-\nu_n^-(-\bar{\tau})\,.
\end{align}
They admit the following Fourier series expansions
\begin{align}
&\nu_n^+(\tau)=\sum_{j=1}^{\infty}a_j^{(n,+)}\cos(2\pi j\tau_1)\sqrt{\tau_2}K_{\ri t_n^+}(2\pi j\tau_2)\,,\\\nonumber
&\nu_n^-(\tau)=\sum_{j=1}^{\infty}a_j^{(n,-)}\sin(2\pi j\tau_1)\sqrt{\tau_2}K_{\ri t_n^-}(2\pi j\tau_2)\,.
\end{align}
The coefficients $a_j^{(n,\pm)}$ and the eigenvalues $t_n^{\pm}$ cannot be written down in a simple form, but can be determined numerically. Their explicit values can be found \emph{e.g} in \cite{terras2013harmonic}.

\section{Properties of integral transformation}
In this appendix, we present detailed proofs for the property of $\mathcal{T}_{\chi}$-transform in the main text.

\vspace{0.3cm} 
\paragraph{Commutativity with modular transformation.}
We first prove the action of $\mathcal{T}_{\chi}$-transform commutes with the action of modular transformation. Recall the $\mathcal{T}_{\chi}$-transform \eqref{eq:defTchi1} is defined as
\begin{align}
\mathcal{T}_{\chi}[f](\tau)=\frac{\chi}{\pi}\int_{\mathbb{H}}\frac{\rd^2\zeta}{(\text{Im}\,\zeta)^2}e^{-\chi\frac{|\tau-\zeta|^2}{\text{Im}\tau\,\text{Im}\,\zeta}}f(\zeta)
\end{align}
Note that $\chi$ does not change under modular transformation. The integral measure is invariant under modular transformation. The kernel has the following property: it is invariant under a simultaneous modular transformation $\gamma$ for $\tau$ and $\zeta$, namely
\begin{align}
\frac{|\gamma\cdot\tau-\gamma\cdot\zeta|^2}{\text{Im}(\gamma\cdot\tau)\text{Im}(\gamma\cdot\zeta)}=\frac{|\tau-\zeta|^2}{\text{Im}\,\tau\text{Im}\,\zeta}\,,
\end{align}
which can be verified by a simple computation. This leads to
\begin{align}
\frac{|\tau-\gamma^{-1}\cdot\zeta|^2}{\text{Im}(\tau)\text{Im}(\gamma^{-1}\cdot\zeta)}=\frac{|\gamma\cdot\tau-\zeta|^2}{\text{Im}(\gamma\cdot\tau)\text{Im}(\zeta)}\,.
\end{align}
Using this property, we have
\begin{align*}
\mathcal{T}_{\chi}[\gamma\cdot f](\tau)=&\,\frac{\chi}{\pi}\int_{\mathbb{H}}\frac{\rd^2\zeta}{(\text{Im}\zeta)^2}e^{-\chi\frac{|\tau-\zeta|^2}{\text{Im}\tau\,\text{Im}\zeta}}f(\gamma\cdot\zeta)\\\nonumber
=&\,\frac{\chi}{\pi}\int_{\mathbb{H}}\frac{\rd^2\zeta'}{(\text{Im}\zeta')^2}e^{-\chi\frac{|\tau-\gamma^{-1}\cdot\zeta'|^2}{\text{Im}(\tau)\text{Im}(\gamma^{-1}\cdot\zeta')}}f(\zeta')\\\nonumber
=&\,\frac{\chi}{\pi}\int_{\mathbb{H}}\frac{\rd^2\zeta'}{(\text{Im}\zeta')^2}e^{-\chi\frac{|\gamma\cdot\tau-\zeta'|^2}{\text{Im}(\gamma\cdot\tau)\text{Im}(\zeta')}}f(\zeta')\\\nonumber
=&\,\mathcal{T}_{\chi}[f](\gamma\cdot\tau)=\gamma\cdot \mathcal{T}_{\lambda}[f](\tau)
\end{align*}
which means that the two operations commute, \emph{i.e.} $[\mathcal{T}_{\chi},\gamma]=0$. 
\vspace{0.3cm} 
\paragraph{Action on Maass forms.}
Now we prove \eqref{eq:actionTMaass}.
For a given $f(\tau)$, its $\mathcal{T}_{\chi}$-transform is uniquely determined and satisfies the initial condition
\begin{align}
\lim_{\lambda\to 0}\mathcal{T}_{\chi}[f]=f(\tau)\,.
\end{align}
Moreover, $\mathcal{T}_{\chi}[f]$ obeys the differential flow equation involving the deformation parameter $\lambda$ \cite{Cardy:2018sdv,Aharony:2018bad}
\begin{align}
\label{eq:flowEq}
\partial_{\lambda}\mathcal{T}_{\chi}[f]=\left[\tau_2\partial_{\tau}\partial_{\bar{\tau}}+\frac{1}{2}\left(\partial_{\tau_2}-\frac{1}{\tau_2} \right)\lambda\partial_{\lambda}\right]\mathcal{T}_{\chi}[f]\,.
\end{align}
We now specialize to $f(\tau)=E_s(\tau)$. Using $\chi=\tau_2/\lambda$ and defining $\mathcal{W}_s=e^{-2\chi}\mathcal{T}_{\chi}[E_s]$, we find that $\mathcal{W}_s$ satisfies
\begin{align}
\label{eq:W}
\left(\Delta_{\tau}+\chi^2\partial_{\chi}^2-4\chi^2 \right)\mathcal{W}_s=0\,.
\end{align}
where $\Delta_{\tau}=-\tau_2^2(\partial_{\tau_1}^2+\partial_{\tau_2}^2)$ is the hyperbolic Laplacian on the Poincaré upper half-plane. Assuming a factorized ansatz $\mathcal{W}_s(\tau,\chi)=E_s(\tau)B_s(\chi)$ and using the eigenvalue equation
\begin{align}
\Delta_{\tau}E_s(\tau)=s(1-s)E_s(\tau)\,,
\end{align}
we obtain the following ordinary differential equation for $B_s(\chi)$
\begin{align}
\chi^2B''_s(\chi)-(4\chi^2+s(s-1))B_s(\chi)=0\,.
\end{align}
The general solution is
\begin{align}
B_s(\chi)=\sqrt{\chi}\left(c_1\,I_{s-\frac{1}{2}}(2\chi)+c_2\,K_{s-\frac{1}{2}}(2\chi) \right)
\end{align}
for $s(1-s)\ne 0$ and 
\begin{align}
B_s(\chi)=
c_1\,e^{2\chi}+c_2\,e^{-2\chi}
\end{align}
for $s(1-s)=0$, where $I_{\nu}$ and $K_{\nu}$ are modified Bessel functions. The limit $\lambda\to 0$ corresponds to $\chi\to\infty$, and the initial condition requires
\begin{align}
\lim_{\chi\to\infty}e^{2\chi}B_s(\chi)=1\,.
\end{align}
This asymptotic behavior fixes $c_1=0$ and $c_2=2/\sqrt{\pi}$. Consequently,
\begin{align}\label{TlambdaEs}
\mathcal{T}_{\chi}[E_s]=e^{2\chi}B_s(\chi)E_s=\frac{2e^{2\chi}}{\sqrt{\pi}}\sqrt{\chi}K_{s-\frac{1}{2}}(2\chi)\,E_s(\tau)\,.
\end{align}
Uniqueness guarantees that this is indeed the $\mathcal{T}_{\lambda}$-transform of $E_s$. The proof for $\nu_n(\tau)$ follows analogously, using the eigenvalue 
\begin{align}
\Delta_{\tau}\nu_n(\tau)=s_n(1-s_n)\nu_n(\tau)=(\tfrac{1}{4}+t_n^2)\nu_n(\tau)
\end{align}
we conclude that
\begin{align}
\mathcal{T}_{\chi}[\nu_n]=\frac{2e^{2\chi}}{\sqrt{\pi}}\sqrt{\chi}K_{\mathrm{i}t_n}(2\chi)\,\nu_n(\tau)\,,\quad n\geq 1.
\end{align}
The $\mathcal{T}_{\chi}$-transform of $E^r_1$ can be computed by taking proper limiting procedure. 
Recall the definition of $E^r_1(\tau)$
\begin{align}
E^r_1(\tau)=\lim_{\epsilon\to 0}\left(E_{1+\epsilon}(\tau)-\frac{3}{\pi\epsilon}\right)\,,
\end{align}
plugging into \eqref{TlambdaEs}, we obtain
\begin{align}
\mathcal{T}_{\chi}[E^r_1]=&\,\lim_{\epsilon\to 0}\left(\mathcal{T}_{\chi}[E_{1+\epsilon}]-\mathcal{T}_{\chi}\left[\frac{3}{\pi\epsilon}\right]\right)\\\nonumber
=&\,\frac{2}{\sqrt{\pi}}e^{2\chi}\left(\frac{3}{\pi}\left.\sqrt{\chi}\partial_{\nu}K_{\nu}(2\chi)\right|_{\nu=1/2}+\frac{\sqrt{\pi}}{2}e^{-2\chi}E^r_1\right)\\\nonumber
=&\,E_1^r(\tau)+\frac{3}{\pi}U(1,1,4\chi)\,,
\end{align}
where in the last line we have exchanged the limit $\epsilon\to 0$ and $\mathcal{T}_{\chi}$-transform. The exchange is valid since expanding $E_s(\tau)B_s(\chi)$ at $s=1+\epsilon$, every $O(\epsilon^n)$ terms satisfies the flow equation and is modular invariant.

\section{Proof of the convergence of eq.~\eqref{eq:TZE}}

In this section, we prove that the infinite sum in \eqref{eq:TZE}, i.e.
\begin{equation}\label{eq:Mtau}
M(\tau)\equiv\sum_{\gamma\in\Gamma/\Gamma_{\infty}}
\left({\rm e}^{2\chi\left(1-\sqrt{1-\frac{2\pi \tc}{12\chi}\text{Im}(\gamma\cdot\tau)} \right)}
 -1-\frac{2\pi \tc}{12}\text{Im}(\gamma\cdot\tau)\right)
\end{equation}
is convergent.
For this purpose, we spell out the range of summation,
\begin{equation}
    \Gamma/\Gamma_\infty = \{(c,d)\in\mathbb{Z}^2 \,|\, \text{GCD}(c,d) = 1\},
\end{equation}
and rewrite the infinite sum as
\begin{equation}
    M(\tau) = \sum_{\gamma:\text{GCD}(c,d)=1}
    \left({\rm e}^{2\chi\left(1-\sqrt{1-\epsilon y_\gamma} \right)}
 -1-\chi \epsilon y_\gamma\right)
\end{equation}
where $\epsilon=\pi \tc/(6\chi)$ and
\begin{equation}
    y_\gamma = \rm{Im}(\gamma\cdot\tau) = \frac{\tau_2}{|c\tau+d|^2}.
\end{equation}
If $(c,d) = (0,\pm 1)$, $y_\gamma = \tau_2$, and when $|c|,|d|$ become greater, $y_\gamma$ becomes progressively small, eventually tending to zero.
In fact, for any value of $\epsilon$, there are only finitely many $\gamma \in \Gamma/\Gamma_\infty$ such that
\begin{equation}
    |\epsilon| y_\gamma  > 1
\end{equation}
or equivalently
\begin{equation}
    |\epsilon|\tau_2 > (c\tau_1+d)^2 + c^2\tau_2^2.
\end{equation}
Let us now split the infinite sum  \eqref{eq:Mtau} to two parts, the part $M_{>}(\tau)$ over the finite subset of $\Gamma/\Gamma_\infty$,
\begin{equation}
    (\Gamma/\Gamma_\infty)_> = \{\gamma \in \Gamma/\Gamma_\infty \,|\, |\epsilon|y_\gamma>1\},
\end{equation}
which is obviously finite,
and the part $M_{<}(\tau)$ over the infinite complement $(\Gamma/\Gamma_\infty)_<$.
The summand of $M_{<}(\tau)$ can be Taylor expanded, yielding
\begin{equation}
    M_{<}(\tau)
    =\sum_{\gamma\in(\Gamma/\Gamma_\infty)_<} 
    \sum_{n=2}^\infty a_n y^n_{\gamma}.
\end{equation}
The power series of $y_\gamma^n$ over $n$ has a radius of convergence $1/|\epsilon|$, and the first few coefficients are
\begin{equation}
    a_2 = \frac{\pi^2\bar{c}^2(2\chi+1)}{144\chi}, \quad a_3 = \frac{\pi ^3 \bar{c}^3 \left(4 \chi ^2+6 \chi +3\right)}{5184 \chi ^2}.
\end{equation}
As the Taylor series is absolutely convergent, we can exchange the order of summation and arrive at
\begin{align}
    M_{<}(\tau) &=\sum_{n=2}^\infty a_n\left(E_n(\tau) - \sum_{\gamma\in(\Gamma/\Gamma_\infty)_>}y_\gamma^n\right)
   \label{eq:MESer}
\end{align}
In the limit $n\rightarrow \infty$, the real-analytic Eisenstein series $E_n(\tau)$ has the asymptotic behavior
\begin{equation}
    \left(E_n(\tau) - \sum_{\gamma\in(\Gamma/\Gamma_\infty)_>}y_\gamma^n\right) \sim \mathcal{O}(y_{\gamma,<}^n),
\end{equation}
where $y_{\gamma,<}$ is the maximum member of $\{y_\gamma \,|\, \gamma\in(\Gamma/\Gamma_\infty)_<\}$, which is by definition smaller than $1/|\epsilon|$, and thus eq.~\eqref{eq:MESer} is also convergent.

\bibliography{yunfeng} 
\bibliographystyle{utphys}
\end{document}